\documentclass[useAMS]{mn2e}
\usepackage{graphics,epsfig,psfig}
\usepackage[dvipsnames]{color}
\usepackage[]{inputenc,amssymb}

\usepackage{graphicx}

\usepackage{color}

\def\:{\ \ \ }
\newcommand{\be}{\begin{eqnarray}}
\newcommand{\ee}{\end{eqnarray}}

\def\diffunits{\rm GeV \, s^{-1}  sr^{-1} cm^{-2}}

\def\etal{{\it et al.} }

\def \be{\begin{equation}}
\def \ee{\end{equation}}
\def \bea{\begin{eqnarray}}
\def \eea{\end{eqnarray}}

\def\simle{\lower 2pt \hbox {$\buildrel < \over {\scriptstyle \sim }$}}
\def\simge{\lower 2pt \hbox {$\buildrel > \over {\scriptstyle \sim }$}}

\def\ARAA{{\it ARA\&A}}
\def\ApJ{{\it ApJ}}
\def\ApJL{{\it ApJ}}
\def\ApJS{{\it ApJS}}
\def\ApP{{\it Astropart. Phys.}}
\def\AA{{\it A\&A}}

\def\AAL{{\it A\& A Lett}}

\def\MNRAS{{\it MNRAS}}
\def\Nature{{\it Nature}}
\def\NewAR{{\it New Astron. Rev.}}

\def\PRD{{\it Phys. Rev.} {D}}
\def\PRL{{\it Phys. Rev. Lett}}
\def\RMP{{\it Rev. Mod. Phys.}}
\def\RPP{{\it Rep. Pro.Phys.}}
\def\Science{{\it Science}}

\title[
Cosmic background from first black holes]{
Cosmic backgrounds due to the formation of the first generation of supermassive black holes}

\author[P. L. Biermann etal]
{\parbox{\textwidth}
{
Peter L. Biermann$^{1,2,3,4}$, Biman B. Nath$^{5}$, Lauren\c{t}iu  I. Caramete$^{6}$, Benjamin C. Harms$^{3}$, Todor Stanev$^{7}$ \& Julia Becker Tjus$^{8}$
}
\vspace{0.4cm}\\
\parbox{\textwidth}{
$^{1}$ MPI for Radioastronomy, Auf dem H\"ugel 69, Bonn, Germany\\
$^{2}$ Dept.\ Physics., Karlsruher Institut f{\"u}r Technologie, Karlsruhe, Germany\\
$^{3}$ Dept. of Phys. \& Astr., Univ. of Alabama, Tuscaloosa, AL, USA\\
$^{4}$ Dept. of Phys. \& Astron., Univ. of Bonn, Germany \\
$^{5}$ Raman Research Institute, Sadashiva Nagar, Bangalore 560080, India\\
$^{6}$ Dept. of Phys., Univ. Bochum, Bochum, Germany; \\
$^{7}$ Bartol Research Inst., Univ. of Delaware, Newark, DE, USA\\
$^{8}$ Institute for Space Sciences, Bucharest, Romania \\
}}

\begin{document}


\maketitle


\begin{abstract}
The statistics of black holes and their masses strongly suggests that their mass distribution has a cutoff towards lower masses near $3 \times  10^{6}$ M$_{\odot}$. This is consistent with a classical formation mechanism from the agglomeration of the first massive stars in the universe. However, when the masses of the stars approach $10^{6}$ M$_{\odot}$, the stars become unstable and collapse, possibly forming the first generation of cosmological black holes. Here we speculate that the claimed detection of an isotropic radio background may constitute evidence of the formation of these first supermassive black holes, since their data are compatible in spectrum and intensity with synchrotron emission from the remnants. The model proposed fulfills all observational conditions for the background, in terms of single-source strength, number of sources, far-infrared and gamma-ray emission. The observed high energy neutrino flux is consistent with our calculations in flux and spectrum. The proposal described in this paper may also explain the early formation and growth of massive bulge-less disk galaxies as derived from the massive, gaseous shell formed during the explosion prior to the formation of a supermassive black hole.
\end{abstract}

\begin{keywords}
Cosmology : first stars -- theory; galaxies : high-redshift -- intergalactic medium ; shock waves ; acceleration of particles
\end{keywords}

\section{Introduction}
Black holes are ubiquitous in the centers of early Hubble type galaxies, and their mass distribution shows evidence for a low mass cut-off near $3 \cdot 10^{6} \, M_{\odot}$ (Greene \etal 2006, 2008; Greene \& Ho 2007a, b; Caramete \& Biermann 2010).  Massive stars readily turn into black holes (Heger \etal 2003, 2005; Woosley \etal 2002), and the agglomeration of massive stars (Spitzer 1969; Sanders 1970; Quinlan \& Shapiro 1990, Portegies Zwart \etal 2004, 2007, 2010; McMillan \etal 2007), perhaps aided by a gravo-thermal collapse (Spitzer 1969, 1987), can turn them into yet more massive stars. However, their powerful winds counteract the increase in mass, and the maximal mass which can be reached from merging is limited to a few hundred solar masses (Yungelson \etal 2008, Crowther \etal 2010).  On the other hand, the wind is driven by radiation interacting with metal ions, thus for zero metal stars there is no such wind, and no mass loss (Heger \etal 2003, Woosley \etal 2002).  It follows that massive, zero-metal stars can indeed reach very high masses in agglomeration.  As the instability refers to that part of the star which is in hydrostatic equilibrium, and the further outer layers of such a rapidly growing supermassive star may still be relaxing towards hydrostatic equilibrium, the fall-back may increase the mass of the final black hole beyond the mass of the hydrostatic mass fraction of the star itself. In such a way the initial black hole mass can possibly become larger than the stellar instability threshold.
An alternate picture is that of the growth of a budding super-massive star by direct accretion of collapsing material (e.g. (e.g., Begelman \etal 2006, Bonoli \etal 2012).  The formation of this super-massive star and then a black hole may be aided by the potential well of a surrounding dark matter clump (see, e.g., Munyaneza \& Biermann 2005, 2006 
; Destri \etal 2012 
).  Massive stars are dominated by radiation pressure, and with increasing zero-age main sequence mass their effective adiabatic gas index approaches the unstable limit of 4/3 (Chandrasekhar 1939). Subtle effects of general relativity push the stars over the limit, causing stars to blow up at a mass approaching $10^{6} \, M_{\odot}$ (Appenzeller \& Fricke 1972a, b).  It follows that the agglomeration of zero-metal stars can readily produce super-massive black holes.  
It is possible in some cosmological models (e.g., Biermann \& Kusenko (2006)) for star formation to occur early ($z \ge 20$), paving the way to an early black hole formation.
Whalen \etal (2013) have recently discussed the possibility of supernovae explosions of massive population III stars which could lead to the formation of the first generation of black holes.
In such cases, massive black holes can grow rather rapidly, by merging with other black holes or by accretion (e.g., Wang \& Biermann 1998, Silk \& Rees 1998, Gergely \& Biermann 2009).  In this way the existence of extremely massive black holes at high redshift can be understood (e.g. Mortlock \etal
2011).  The formation of super-massive black holes at very high redshift may have consequences in galaxy evolution and cosmology (e.g. Kormendy \etal 2010, 2011, Kormendy \& Bender 2011, Conselice \etal 2011, Biermann \& Harms 2012, 2013a, b, Buitrago \etal 2013, Conselice \etal 2013).

This line of reasoning suggests that it is worth exploring 
possible observational tests in order  to confirm or refute such a picture.

The recent claim of the detection of an isotropic radio background unrelated to any known population of galaxies (Kogut \etal 2011, Fixsen \etal 2011, Seiffert \etal 2011, Condon \etal
2012, Formengo \etal 2014; but see also Subrahmanyan \& Cowsik 2013)) raises the possibility that the explosions of these super-massive stars which give rise to the first generation of super-massive black holes might be considered as being similar to a supernova explosion with the concomitant acceleration of particles producing radio emission.  We work out the strength of this radio emission.  We do not question here whether this is the only possible explanation, as clearly the thick cosmic ray disk in our own Galaxy can also produce strong background emission (Sun \etal 2008; Everett \etal 2010), but it seems unlikely to be able to give the spectrum derived by Fixsen \etal (2011).  However, the strength of the spectral constraint depends on the  error estimate.  As a test of our model, we consider the conditions derived by Condon \etal (2012), and show that their conditions can all be met; these include the strength of each source, the number of sources, and the absence or weakness of far-infrared emission.  The flux density determined for the background, which was not explained by known source populations by Condon \etal (2012), can be determined using the spectrum obtained by Fixsen \etal (2011). The observed flux density at 3.02 GHz corresponds at 1 GHz to $10^{-18.5} \, {\rm erg \, cm^{-2} \, s^{-1} \, Hz^{-1} \, sr^{-1}}$ with an error of about 16 percent. 
Earlier attempts to interpret this radio background were done by Singal et al. (2010), Meiksin \& Whalen (2013), and Holder (2014). They noted some of the same difficulties emphasized by Condon et al. (2012).

\section{Nonthermal radio background}
Assuming the scenario as outlined above, we can consider the further evolution of such black holes, formed with a mass near $3 \cdot 10^{6} \, M_{\odot}$, as suggested by the mass distribution of black holes (Caramete \& Biermann 2010, and references therein 
).  
In the following analysis we focus on the non-thermal radio emission from the remnants formed, after the super-massive stars explode.  As the reference redshift for the formation epoch of the first population of super-massive black holes we cautiously adopt 20 (see, e.g., Kogut \etal 2003) and write $1+z = z_{1.3} (1+20)$, so higher redshifts could be allowed for if required. 

The scenario which we use to produce the very first generation of super-massive back holes was first worked out by Spitzer (1969) and Sanders (1970), combined with the work by Appenzeller \& Fricke (1972a, b).  That was long before we knew how ubiquitous super-massive black holes are.  More recently this question has been explored by many, e.g. by Begelman \etal (2006), Bellovary \etal (2011), and Zinn \etal (2011).  The consequences for ultra high energy neutrinos have been worked out by Berezinsky \& Blasi (2012). In this paper we concentrate on lower energies of both neutrinos and gamma-ray photons.

When a supermassive star with a mass of about or larger than $3 \times 10^{6} \, M_{\odot}$ explodes, 
rather than a star of order $30 \, M_{\odot}$ in zero-age mass (Woosley \etal 2002) explodes producing a black hole, we assume that the total energy that can be transferred to baryonic material scales with the final black hole mass.  Assuming that gamma ray bursts leave massive stellar mass scale black holes behind, of order $5 \; M_{\odot}$ and explode with about $10^{51}$ to $10^{52}$ ergs (see, e.g. Cox 1972, Nakamura \etal 1999, H{\"o}flich \etal 1999, Pugliese \etal
 2000), we  use the lower number as a conservative reference.  We will use an efficiency of turning $M_{BH} c^2$ into electrodynamic energy of about $10^{-4}$.  This corresponds to $10^{56.8}$ ergs, and so as an approximation we will adopt $10^{57} E_{57}$ erg as a reference. 

The explosion of a super-massive star is assumed to be a scaled-up version of an ordinary supernova explosion and to be moving through a medium of density $n_0 (1 + z)^{3}$ with $n_0\sim 2 \times 10^{-7}  \, {\rm cm^{-3}} $ as the particle density today (e.g., 
Ade \etal 2013; PLANCK 2013 XVI). In the corresponding Sedov-Taylor phase of these cosmological blast-waves (Voit 1996; McKee \& Ostriker 1988), in which the swept-up mass of the shell is larger than the ejecta mass, the radius for the blast-wave originating at a redshift $z_0$ is given by,
\begin{eqnarray}
R&&={\left(\frac{\zeta \, E}{2 n_0 m_{H,He} (1+z_0)^{3}}\right)}^{1/5} \, (\Delta t)^{2/5} \nonumber\\
&& \sim \, 10^{22.76} \, E_{57} ^{1/5} \, z_{1.3}^{-3/5} \, {\{\Delta t\}_{15}}^{2/5} \, {\rm cm}\,,
\end{eqnarray}
where $\zeta=2.025$, $m_{H,He} \simeq 10^{-23.7}  \, {\rm g}$ is the average mass of nuclei for a primordial mixture of hydrogen and helium, and $\Delta t$ is the time elapsed since the origin of the blast-wave ($\{\Delta t\}_{15}$ is $\Delta t$ in units of $10^{15}$ s). Voit (1996) showed that  the blast-wave radius reaches an asymptotic value at large time scales, but the blast-waves would dissipate their energy after the speed decreases below $\sim 10$ km s$^{-1}$, when the post-shock temperature is of order $10^5$ K and the cooling rate is large. The blast-wave speed is given by,
\begin{eqnarray}
\dot{R}&&\sim (2/5){ \left(\frac{\zeta \, E}{2 n_0 m_{H,He} (1+z_0)^{3}}\right)}^{1/5} \, (\Delta t)^{-3/5} \nonumber\\
&& \sim  10^{2.4} \, E_{57}^{1/5} \, z_{1.3}^{-3/5} \, {\{\Delta t\}_{15}}^{-3/5} \, {\rm km} \, {\rm s}^{-1} \,.
\end{eqnarray}
We first notice that the blast-wave becomes non-relativistic at a time $\{\Delta t\}_{15} \sim 9 \times 10^{-6} \, E_{57}^{1/3} \, z_{1.3}^{-1}$.  However, what is  more important is that the environmental mass encountered becomes of similar order to the mass of the exploding star only at relatively large radius, corresponding to about $\{\Delta t\}_{15} \, \sim \, 10^{-2}$.  This time interval corresponds in terms of a radial scale to the fraction ${\{\Delta t\}_{15}}^{2/5} \, \sim 10^{-0.8}$, only about one order of magnitude below the scale introduced above.  We will use this scale as our  limit for the time up to which the blastwave will accelerate cosmic rays. 

Also, the radiative cooling phase begins at a time scale of 
$t_{rad}\sim 10^{15.8}  \, E_{57}^{1/3} \, z_{1.3} \, {\rm s}.$
Interestingly, the inverse Compton cooling time-scale of hot electrons in cosmological shock waves against the cosmic microwave background is 
also of the same order (Tegmark, Silk, Evrard 1993), $t_{IC}\sim 1.78 \times 10^{15}  \,
z_{1.3}^{-4} \, {\rm s}$, independent of the blast-wave energy. For comparison, the age of the universe is $6.17\times 10^{15} \, z_{1.3}^{-3/2}$ s, similar to or larger than the time scales considered above. 
Using the cooling limit implies an activity redshift interval of 
$\Delta z \; = \; 10^{1.15} E_{57}^{1/3} \; z_{1.3}^{7/2} \, .$ Of course $\Delta z$ cannot exceed $1+z$ itself, so using this expression is limited to redshifts close to 20; for any significantly larger redshift $(2/3)(1+z)$ is an approximate limit.  Correspondingly, the local Hubble time is a stronger limit in case the Hubble time is shorter than the cooling time.  The Hubble time at high redshift is
\begin{equation}
t_{H}\sim 10^{16} \,  z_{1.3}^{-3/2} \, {\rm s} \,.
\end{equation}
Since we propose to cover a range of redshifts, we will use this limit in the following. In other words, ${\{\Delta t\}_{15}} \approx 10 \, z_{1.3}^{-3/2}$.
We can therefore assume that the blast-waves reach a distance given 
by 
\begin{equation}
R_{lim}\sim 10^{23.16} \, E_{57} ^{1/5} \, z_{1.3}^{-6/5} \, {\rm cm} \,.
\end{equation}
Another limit is the radius at which the various maximal spheres touch each other, given by
\begin{equation}
R_{space}\sim 10^{23.2} \, N_{BH,0}^{-1/3} \, z_{1.3}^{-1} \, {\rm cm} \,.
\end{equation}
which is of similar size, but somewhat larger, especially at larger redshifts, using as reference for the original black hole (comoving) space density $N_{BH,0} \; = \, 1$ Mpc$^{-3}$ and for the explosion energy $E_{57} \; = \; 1$ .  

Assuming a shell thickness $\Delta R=R/12$ in the limit of a strong shock with density jump of a factor of 4, the volume of the shell is $\sim R^3$. The magnetic field and the particle content in the shell can then be written as,
\begin{equation}
\frac{B^{2}}{8 \pi} \, R^{3} =  \eta_{B} \, E \,; \qquad 
\frac{m_e c^2}{p - 2} \, C\, R^{3}\; = \; \eta_{CR,e} \, E \, , 
\end{equation}
where $p$ is the spectral index of the particle energy distribution, written as $C \gamma_e^{-p} \, d \gamma_e$, and $\gamma_e$ is the Lorentz factor of the cosmic ray electrons; $C$ is defined as the amplitude of the energetic electron spectrum.  We assume the fiducial value of the final 
fraction of energy transferred to electrons to be $\eta_{CR,e}=0.1 \, \eta_{CR,e, -1}$, as well as $\eta_{B} = 0.1 \eta_{B,-1}$ for the fraction of energy transferred via instabilities to magnetic fields (Weibel 1959, Lucek \& Bell 2000; Bell \& Lucek 2001). We use as fiducial values for the energy transfer 10 \% in each case.  We assume that there are only negligible magnetic fields already present. Also, we adopt a spectral index $p=2.2$ to match the radio data.
Then we have, 
\begin{eqnarray}
B &\approx & 10^{-5.44}  \, \eta_{B,-1}^{1/2} \, E_{57}^{1/5} \, z_{1.3}^{9/10} \,
\{\Delta t\}_{15}^{-3/5}\, {\rm Gau{\ss}} \, 
\nonumber\\
C&\approx &10^{-6.9} \, \eta_{CR,e,-1} \, E_{57}^{2/5}\, z_{1.3}^{9/5} \, \{\Delta t\}_{15}^{-6/5}\, {\rm cm}^{-3}\,. 
\end{eqnarray} 

Finally, the radio luminosity per frequency can be written as
\begin{eqnarray}
L_{\nu} &=& 10^{29.99} \, \eta_{B,-1}^{0.80} \, \eta_{CR,e,-1}^{+1} \, E_{57}^{1.32} z_{1.3}^{1.84} {\nu}_{9.0}^{-0.60} \nonumber\\
&& \times  \{\Delta t\}_{15}^{-0.96} \, {\rm erg} \,{\rm s}^{-1}  \,{\rm Hz}^{-1} \,. 
\end{eqnarray}
\noindent including the spectral k-correction $(1+z)^{1-0.6}$, where $0.6$ is the radio spectral index.  Inserting the limiting radius derived above, this can be rewritten as,
\begin{equation}
L_{\nu}  = 10^{28.99} \, \eta_{B,-1}^{0.80} \, \eta_{CR,e,-1}^{+1} \, E_{57}^{1.32} z_{1.3}^{3.34} {\nu}_{9.0}^{-0.60} \, {\rm erg} \, {\rm s}^{-1}  {\rm Hz}^{-1} \,. 
\end{equation}
It is straightforward to verify that the remnant is not optically thick to synchrotron self-absorption.  However, since the emission varies with time as $\sim t^{-1}$, we must take an average over various evolutionary stages of such explosion bubbles and obtain an extra factor from the logarithm of the ratio of the longest to the shortest time, $\ln {t_{max}}/{t_{min}}$. Identifying naively the minimal radius as the one 
where the motion becomes adiabatic we obtain a factor  $\sim 7$, giving
\begin{eqnarray}
L_{\nu}  &= &10^{29.82} \, \eta_{B,-1}^{0.80} \, \eta_{CR,e,-1}^{+1} \, E_{57}^{1.32} \, z_{1.3}^{3.34} \, {\nu}_{9.0}^{-0.60} \, 
\nonumber\\
&& {\rm erg } \, {\rm s}^{-1} \, {\rm Hz}^{-1} \, . 
\end{eqnarray}

The radio background can be written as
\begin{equation}
F_{\nu} = N_{BH,0} \, \frac{c \, r(z)^{2}}{H(z)} \, \frac{L_{\nu}}{4 \pi d_{L}^{2}} \, \Delta z \, ,
\label{bgeq}
\end{equation}
with the units of ${\rm erg} \, {\rm  s}^{-1} \, {\rm Hz}^{-1} \, {\rm cm}^{-2} \, {\rm sr}^{-1}$, and where $N_{BH,0}$ is the comoving number density of these explosions; $L_{\nu} $ is the radio luminosity per frequency of a single explosive event; $r(z)$ is the comoving distance and ${d^2 V \over dz d\Omega}={cr(z)^2 \over H(z)}$ is the comoving volume element per unit redshift and solid angle. Also, $\Delta z=(2/3) (1+z) $ is the redshift interval for which the radio emission is maintained. By definition, we have $d_L(z) \, =\, r(z) \, (1+z)$, with the asymptotic limit of $r(z) \, \rightarrow \, 10^{4.165}$ Mpc at high redshift, so we can write,
\begin{eqnarray}
F_{\nu}  &\approx & 10^{-19.8} \, N_{BH, 0, 0} \, \eta_{B,-1}^{0.80} \, \eta_{CR,e,-1}^{+1} \, E_{57}^{1.32} \,  z_{1.3}^{+0.84} \, {\nu}_{9.0}^{-0.60} \nonumber\\
&& \, {\rm erg} \, {\rm s}^{-1} \, {\rm Hz}^{-1} \, {\rm cm}^{-2} \, {\rm sr}^{-1}\, .
\end{eqnarray}
Here $N_{BH,0}=1 \, N_{BH, 0, 0}$ Mpc$^{-3}$, and we have used a Hubble constant of $h=0.7$ (see Planck 2013 XVI).


After taking into account the contributions of known sources and using the observed spectrum to interpolate to $10^{9} \, {\rm Hz}$ (Kogut \etal 2011, Fixsen \etal 2011, Seiffert \etal 2011, Condon \etal 2012) the observations suggest a flux density of $10^{-18.5} \; {\rm erg}  \, {\rm s^{-1} \, Hz^{-1} \, cm^{-2} \, sr^{-1}}$. 
%
%
This implies that in order to match this observation at the GHz level we have the condition:
\begin{equation}
10^{+1.3} \; = \; N_{BH, 0, 0} \, \eta_{B,-1}^{0.80} \, \eta_{CR,e,-1}^{+1} \, E_{57}^{1.32}  \, z_{1.3}^{+0.84} \, . 
\end{equation}

This constraint has large uncertainties.   Caramete \& Biermann (2010) gave an integral density today of {$10^{-2.2 \pm 0.4} \, {\rm Mpc^{-3}} \, (M_{BH}/(10^7 M_{\odot}))^{-1}$}, so at our nominal black hole mass of $3 \cdot 10^{6} \, M_{\odot}$ this is $N_{BH, 0, 0} = 10^{-1.7 \pm 0.4}$.  We have to note that this number for today's density of black holes may be a very serious under-estimate for the original density, if black holes grow by merging more than by accretion. Assuming merging is the more important process for the growth of black holes increases $N_{BH, 0, 0}$ by a factor of order $10^{0.9}$ (Caramete \& Biermann 2010).  Additionally, there are systematics, since the sample of galaxies chosen in Caramete \& Biermann (2010) was tightly constrained and missed many types of galaxies with known central black holes, for which defining a complete sample with clear properties is uncertain. This may well account for another factor of 2 or 3 uncertainty that the derived black hole density is too low.  At the 1-$\sigma$ level of statistical errors this may add up to a factor of $10^{0.4 + 0.9 + 0.5} = 10^{1.8}$ for the ratio of the original density to today's density.  Thus the original density may approach $1 \, {\rm Mpc^{-3}}$ or perhaps even exceed it; this is the density which we use above as a reference. Since black holes in galaxies in their quiescent stage are easily overlooked in observations (e.g. Stern \etal
2012), the errors may be even larger, allowing a large original black hole density even with dominant growth by accretion. This number is already implicit in the expression above.  
It is also possible that the redshift can be as large as $z\sim 70$ in some cosmological models (e.g, Biermann \& Kusenko (2006)), so that  the parameter $z_{1.3}^{0.84}$ might be $10^{0.3}$. The explosion energy could easily be higher or lower.  However, we will show below that it is constrained by observations in the context of the model presented here.  The two efficiencies of turning energy into magnetic fields or cosmic ray electrons are conservative guesses.  Finally, the uncertainty in the radio spectral index also translates to an uncertainty in the flux density. Decreasing the spectral index by one sigma of the observations (0.036) increases the predicted flux density by a factor of about 3.  To summarize, the largest uncertainty is the original black hole density; it may account for most of the entire factor we require here.  Based on Condon \etal  (2012)  we will show below that this is rather likely in the context of the model approach used here.

One other uncertainty is whether this simple explosion picture is correct. As an alternative one could also consider the steady feeding of a remnant bubble from accretion or spin-down of central compact objects such as neutron stars or black holes (see, e.g., SS433 in W50, Downes \etal
1986; Weiler \etal 2002).  Working out the final single-source luminosity and integrated flux density yields numbers and expressions not significantly different from those given above. The main differences are the time-evolution of the source and the redshift dependence, since different cut-off arguments must be introduced.  Obviously, a picture can also be developed in which the mass shell ejected by the super-massive star is a relatively large fraction of the star's mass, and then the initial remnant evolution is free expansion, with a constant velocity of the shock front throughout this phase.  Such an evolution would then strongly depend on the ejected mass fraction.


\section{Observational checks}
\subsection{The radio background}
A first test is the flux density of sources presented in Condon \etal (2012). Condon \etal
(2012) were able to set an upper limit on the strength of individual sources to be $< 30$ nJy.  
The model above obeys this limit, giving as a function of time (in erg cm$^{-2}$ s$^{-1}$ Hz$^{-1}$)
\begin{equation}
S_{\nu}  = 10^{-31.0} \, \eta_{B,-1}^{0.80} \, \eta_{CR,e,-1}^{+1} \, E_{57}^{1.32} \, z_{1.3}^{-0.16} \, \{\Delta t\}_{15}^{-1} \, {\nu}_{9.0}^{-0.60} \,. 
\end{equation}

We note that the earliest time with a very brief duration at which our approximations hold increases the flux density by perhaps two orders of magnitude, reaching about 1 $\mu$Jy, which is outside the
range of data in Fig 1 of Condon \etal
(2012).  Our approximations put the flux density at the confusion limit of these observations, but their number density on the sky is reduced for this short phase of the evolution.

The average value of this is (in erg cm$^{-2}$ s$^{-1}$ Hz$^{-1}$; after taking into account the above mentioned limit  ${\{\Delta t\}_{15}}$ and the factor of $\ln t_{max}/t_{min}\sim 7$)
\begin{equation}
S_{\nu} \; = \; 10^{-31.2} \, \eta_{B,-1}^{0.80} \, \eta_{CR,e,-1}^{+1} \, E_{57}^{1.32} \, z_{1.3}^{+1.34} \,  {\nu}_{9.0}^{-0.60} \,, 
\end{equation}
which may need to be raised to allow a match of the radio flux density of the radio background.  This match might be accomplished by allowing the original black hole density $N_{BH,0,0}$ to be larger, and/or one of the other parameters to be larger, since all enter with positive exponents. This gives the factor of $10^{1.3}$ derived above for matching the radio background.  Therefore this flux density is likely of order 20 nJy, with all the uncertainties noted earlier. This is still well below the required limits (Condon \etal 2012).  A corollary is that the far-infrared emission has to be negligible.  This condition is also fulfilled, since in the model proposed the massive agglomerating stars all coalesce and blow up before any heavy elements have been formed in significant quantities.  However, we will note another test on the far-infrared emission below.

A second test is the number of sources on the sky and their possible angular overlap.  The number of visible sources per solid angle can be written as $N_{obs} \; = \; N_{BH, 0} \, \frac{c \, r(z)^{2}}{H(z)} \, \Delta z$.  Using again a  fraction of the expansion time scale of the universe $\Delta z \, = \, (2/3) \, (1+z)$ we obtain the total number of sources per solid angle to be $10^{9.8} \, N_{BH,0,0} \, \, z_{1.3}^{-1/2}$.  This number does not match the requirements derived by Condon et al. (2012) of $10^{11.8}$ sr$^{-1}$. However, using this limit from Condon et al. (2012) is problematic, since the sources considered here are not point sources, but actually somewhat larger than the beam used by Condon et al. (2012). We discuss the beam smearing in the following. 
Condon \etal (2012) 
used a beam of 8 arc sec resolution, while the angular extent of the remnants discussed here is 
\begin{equation}
\theta \; = \; 10^{-4.2} \, E_{57}^{1/5} \, z_{1.3}^{-1/5} \, {\rm rad} \, ,
\end{equation}
which is of order 12 arc sec radius, and they overlap considerably.  It is not clear how a large number of overlapping, slightly extended sources would be detectable in the type of analysis given by Condon \etal (2012).  Requiring the diameter of the sources to be less than a resolution element implies a fraction of $10^{-5}$ of the full time of evolution. The flux density is correspondingly higher by $10^{+4.2}$ (taking out the factor of 7 from the averaging), but with the numbers reduced by the factor of $10^{-5}$. This is close to the limits of the Condon \etal
 (2012) analysis, but still satisfies them.

 We can ask how many sources we should have on the sky which are in the relativistic growing phase of early expansion, whose duration is given by:
\be
\{\Delta t\}_{15} \simeq 10^{-5} E_{57}^{1/3} \, z_{1.3}^{-1}
\ee
from our discussion of Eqn 2. Expanding the time-redshift dependence for high redshift gives
\be
\Delta t \; = \; \tau_{H} \frac{3}{2} (1+z)^{-5/2} \, \Delta z \,.
\ee
Equating these two time scales gives us
\be
\Delta z \simeq 10^{-4.6} E_{57}^{1/3} \, z_{1.3}^{+3/2} \,.
\ee
Inserting this into the number per steradian on the sky gives then for the sources still in the relativistic stage $10^{5.9} \, N_{BH,0,0}$, 
independent of redshift. It is not certain how bright the sources would be at this early stage, but almost certainly still in the growing stage for radio emission.

Since the number of sources per angular resolution element is large, of order $10^{3.3} N_{BH, 0,0} z_{1.3}^{-1/2}$, considerable smearing will occur. Therefore along any given line of sight the number of sources is even larger (see below), so that just by Poisson noise the fractional residual flux variations will be small, of order $\simeq$ $10^{-1}$. 
 This implies that the equivalent source number density should increase by a factor of $10^2$, thereby taking the earlier estimate of $10^{9.8}$ sr$^{-1}$ to $10^{11.8}$ sr$^{-1}$, matching the requirement of Condon \etal (2012). Only the very early brief phases of the evolution (the luminosity runs as $t^{-1}$) are reaching close to the limit of current surveys, as noted. This answers the questions raised by Singal et al. (2010) and Meiksin \& Whalen (2013) about the smoothness of the background. We note in passing that Holder's (2014) discussion provides limits for redshifts below about 5, whereas we consider here redshifts beyond about 20, since our model for the formation of supermassive black holes works only in a near-zero heavy element abundance environment.

A third test is whether we can reproduce the spectrum determined by Fixsen \etal
(2011) implying a particle spectrum of $E^{-2.198 \pm 0.072}$.  The explosion is into a medium of constant density, but no pre-existing magnetic field.  In diffusive shock acceleration (Fermi 1949, 1954, for a review see Drury 1983), as applied to exploding stars, particles are considered to be scattered back and forth across a spherical shock region, gaining energy from the compressed system due to both sides of a shock, losing energy adiabatically from an expansion of the system, and also getting eliminated from the system. In the standard limit for a strong plane-parallel shock in a gas of adiabatic gas constant $5/3$ this gives for relativistic particles of energy $E$ a  spectrum of $E^{-2}$.  In the moving shock frame any magnetic field produces an electric field, and particles experience a drift from the combined action of the magnetic and electric fields, giving some additional energy gain (Jokipii \etal 1977, Jokipii 1987).  This drift derives from both gradients and curvature of the magnetic field, since the expansion is spherical.  A magnetized shock moving into a region without any pre-existing magnetic field thus has curvature from the turbulent motions, and also a gradient from bringing in a new magnetic field (Biermann 1993, Biermann \& Cassinelli 1993, Biermann \& Strom 1993).  This strong gradient doubles the drift energy gain contribution compared to a shock moving into a given magnetized region, which occurs in a normal supernova-explosion in the interstellar medium.  In terms of the language of paper CR-III (Biermann \& Strom 1993) this implies that in Eq. 15 of that paper $x - 1 = 1/3$ instead of $x -1 = 1/6$.  Substituting the first value for $x$ into Eqs. 22 and 26 gives a particle spectrum of $E^{-2.24 \pm 0.04}$,  yielding a radio spectrum of ${\nu}^{-0.62 \pm 0.02}$.  This spectrum is to be compared with the measured radio spectrum of ${\nu}^{-0.599 \pm 0.036}$ (Fixsen \etal 2011).  

A fourth test involves the range of the spectrum, which has been observed to 10 GHz, but which may go  higher in frequency.  Using the temporal dependencies of the magnetic field derived, the calculated value of the maximum emission frequency can be shown to be higher than the observed 10 GHz, and thus the maximum observed radio frequency does not produce a serious constraint on parameters.

Next we work out the predicted neutrino and gamma-ray spectra from hadronic interactions.


\subsection{The diffuse neutrino and $\gamma$-ray background}

\subsubsection{Normalization at one source}
A fraction, $0.1 \, \eta_{CR, -1}$, of the energy of the explosions described above goes into cosmic rays,
\begin{equation}
E_{CR}\approx 10^{56}\,\eta_{CR,-1}\,E_{57}\,{\rm erg}\,,
\end{equation}
with $E_{SN}=E_{57}\cdot 10^{57} {\rm erg}$ as the explosion energy, and
$\eta_{CR}=0.1\cdot \eta_{CR,-1}$ again as the fraction transferred to CRs.
It is further assumed that the CR spectrum follows a power-law with index
$p$ and an exponential cutoff at the maximum energy $E_{\max}$,
\begin{equation}
\frac{dN}{dE}=A_p\cdot E^{-p}\cdot \exp(-E/E_{\max})
\end{equation}
with $p$ again later set to $2.2$ and $E_{\max} = 10$~PeV, compatible with the space available for the Larmor motion. 
The units of the spectrum are particles per TeV.
The spectrum is then normalized via the total energy at the source,
\begin{eqnarray}
E_{CR}&=&\int \frac{dN}{dE}E\,dE\approx A_p\,\int_{E_{\min}}^{E_{\max}}
E^{-p+1}\,dE \\
&=&A_{p}\,\frac{1}{p-2}\,\left(E_{\min}^{-p+2}-E_{\max}^{-p+2}\right)\,,
\end{eqnarray}
for $p>2$, where we replaced the exponential with a cut in the integral, and worked everything out in the source frame.  Here we assume that the limiting energies $E_{min}$ and $E_{max}$ are set such that most of the emission is encompassed.  The range of energies observed, from $E_{1, obs}$ to $E_{2, obs}$, including the redshift factor, must be contained in the source energy range, so that
\begin{equation}
E_{min} \, \simle \, (1+z) E_{1, obs} < (1+z) E_{2, obs} \simle \, E_{max} \,.
\end{equation}
It follows that,
\begin{eqnarray}
A_{p}&=&E_{CR}\cdot (p-2)\,\left(E_{\min}^{-p+2}-E_{\max}^{-p+2}\right)^{-1} \nonumber\\
&=&6\times10^{55}(p-2)\,\eta_{CR, -1} \, E_{57} \, \nonumber\\
&& \times 
\Bigl [ \Bigl ( \frac{E_{\min}}{TeV}\Bigr )^{-p+2}-\Bigl ( \frac{E_{\max}}{TeV}\Bigr )^{-p+2}\Bigr ]^{-1}\,{\rm TeV}^{-1} \,.
\end{eqnarray}

We can estimate the maximum energy of these neutrinos from the maximum energy of protons, which in turn is essentially given by the spatial limitations for the Larmor motion in the expanding shell behind the shock.  Using an estimate of maximum neutrino energy as 1/20 of maximum proton energy based on pion decay, and red-shifting the energy down to the observer frame yields
\be
E_{neutr, max} \; = \; 10^{16.0}  \, \eta_{B,-1}^{1/2} \, E_{57}^{2/5} \, z_{1.3}^{-2/5} \,  {\rm eV}\,,
\label{emax_cr}
\ee
with $\eta_{B,-1}$ again the efficiency with which blast wave energy is transformed into magnetic fields, in units of 0.1.

Following Kelner \etal (2006), the neutrino flux can be determined as,
\begin{equation}
\frac{dN_{\nu}}{dE_{\nu}}=n_H\cdot c\cdot \int \sigma_{pp}
\frac{dN}{dE}\left(\frac{E_{\nu}}{x}\right)\cdot f_{\nu}\left(\frac{E_{\nu}}{x}\right)  \frac{dx}{x} \,.
\label{Fluxeq}
\end{equation}
Here,  $x$ is the fraction of energy transferred from the CR to the neutrinos and  $f_{\nu}$ is the probability distribution for one interaction. The inelastic cross section $\sigma_{pp}$ increases logarithmically with
energy.  In the expressions above the physics model is worked out in the source frame.
Units in this case are particles per TeV and per second.  Also, $n_H$ is proportional to  $(1+z)^3$, so the time-integrated neutrino emission scales just with the total energy deposited in cosmic rays, {\it i.e.} with $(1+z)^3$.  The time period of emission varies as $ \sim \, (1+z)$, as argued above.  Any energy flux $E_{\nu}^2 \frac{dN_{\nu}}{dE_{\nu}}$ evaluated in the observer frame is proportional to $(1+z)^{-2}$ for large redshifts for an $E^{-2}$ spectrum, and assuming a complete spectral coverage. Given a spectrum steeper than $E^{-2}$ the energy flux above some minimum energy defined by the observer introduces an additional  spectral correction of $(1+z)^{-0.2}$, which is about a factor of 2 for the nominal redshift of 20. Using Eq. \ref{bgeq} then gives a total redshift dependence of $(1+z)^{+0.8}$ for the neutrino background energy flux.  The strongest effect here derives from the inverse time-scale as a function of redshift, which scales as $(1+z)^{3/2}$, and so cancels the $z$-dependence of $H(z)$, while the density $n_H$ wins by one over $d_L^2$.  

\begin{figure}
\centering{
\includegraphics[width=\linewidth]{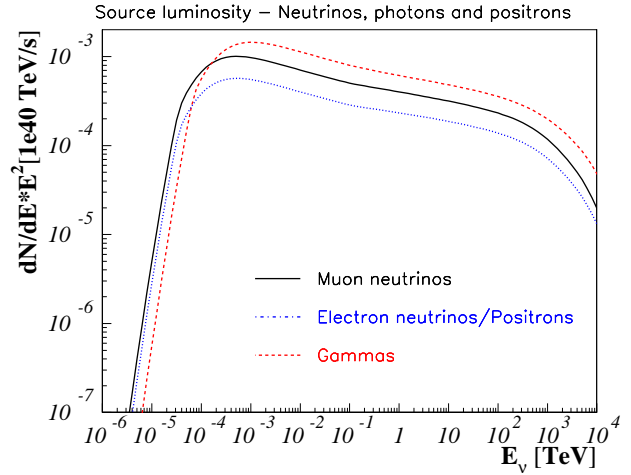}
\caption{Neutrinos and photons at a source. The black line represents muon
  neutrinos; the blue dotted line is for electron neutrinos and the red dashed line is photons.\label{nus_source:fig}}
}
\end{figure}

The photon flux from hadronic interactions is determined in a similar
way (see Kelner \etal (2006)).
The result for the flux from one source at the source redshift is shown in Fig.\ \ref{nus_source:fig}.

\subsubsection{The diffuse flux}
The diffuse neutrino (or photon) flux can be calculated assuming a comoving black hole
density of $N_{\rm BH}
=1$~Mpc$^{-3}$, which is assumed to be constant since a formation epoch at redshift $z_0$. Here, again we use a redshift of 20 as reference, $z_0\approx 20$. Including oscillations, we add the locally produced
muon- and electron-neutrino fluxes and divide by three,
\begin{equation}
\Phi =
\frac{1}{3}\left(\frac{dN_{\nu_{\mu}}}{dE_{\nu_{\mu}}}+\frac{dN_{\nu_{e}}}{dE_{\nu_{e}}}\right) \,
N_{BH} 
\frac{dV}{dz}\,\frac{1}{4\pi\,d_{L}^2}\, \Delta z \,.
\end{equation}
The units here are number per unit area, per energy interval, per steradian, and per time interval.  Using $dV/dz\approx 10^{12.24}$~Mpc$^3\cdot (1+z)^{-3/2}$ and
$d_{L} \, \simeq \, 10^{4.1} \, (1+z)$ Mpc in the high redshift limit, it follows that
\begin{eqnarray}
N_{BH} 
\frac{dV}{dz}\,\frac{1}{4\pi\,d_{L}^2}=10^{-46.6} \, N_{BH,0,0} \, (1+z)^{-7/2}  \, {\rm
  cm}^{-2} {\rm sr}^{-1} \,.
\end{eqnarray}

\begin{figure}
\centering{
\includegraphics[width=\linewidth]{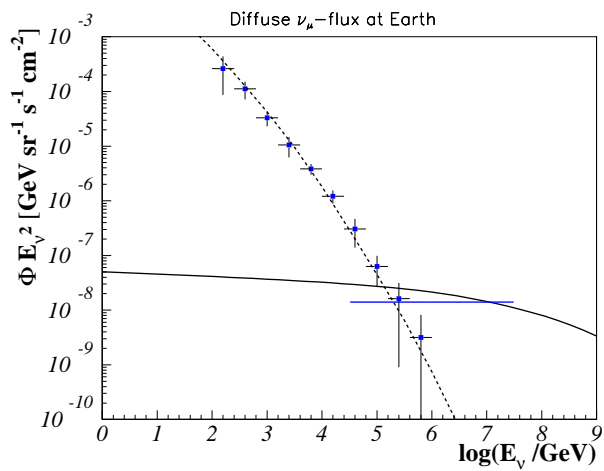}
\caption{Neutrino flux at Earth\label{earth}. Data points show the
  atmospheric neutrino spectrum as measured by IceCube (Ruhe \etal
2013), the dashed line represents the angle-averaged prediction for
  the conventional flux of atmospheric neutrinos (Volkova 1980). The
  horizontal line shows the most recent limit on the diffuse
  extraterrestrial neutrino flux from muon- and antimuon-neutrinos
  (Schukraft \etal 2013).}
}
\end{figure}

The diffuse neutrino differential flux is therefore given as
\begin{eqnarray}
\Phi &= &10^{-46.6}\,{\rm
  cm}^{-2}\,{\rm sr}^{-1} \,{\rm TeV}^{-1} \, {\rm s}^{-1} \nonumber\\
  && \times
\frac{1}{3}\left(\frac{dN_{\nu_{\mu}}}{dE_{\nu_{\mu}}}+\frac{dN_{\nu_{e}}}{dE_{\nu_{e}}}\right)\,,
\end{eqnarray}
using a redshift interval of $(2/3) (1+z)$, as derived earlier, approximately appropriate for a redshift of order 20 and above.  As shown above this diffuse flux as number per energy interval, area, solid angle and time interval scales with redshift as $(1+z)^{-1.2}$. In terms of the integrated energy flux, the scaling with redshift is  modified  to $(1+z)^{+0.8}$. This diffuse integrated energy flux is then
\begin{eqnarray}
\Phi \, E_{\nu}^2 
&=&  10^{-7.5} \, \rm GeV \, {\rm
  cm}^{-2}\,{\rm sr}^{-1} 
  \, {\rm s}^{-1} \, \nonumber\\
  && \times N_{BH, 0, 0} \, E_{57} \, \eta_{CR, -1} \, z_{1.3}^{+0.8} \, \Delta z \,.
 \end{eqnarray}
  This diffuse neutrino flux at Earth is shown in Fig.\ \ref{earth}, together with the atmospheric neutrino spectrum as measured with
IceCube (Ruhe \etal 2013) and the most recent limit on the diffuse
neutrino flux (Schukraft \etal 2013). Latest results (Aartsen \etal 2013) indicate a signal from extraterrestrial sources at a flux
level with a spectrum significantly flatter than the flux of
atmospheric neutrinos. As this prediction is approximately at the
sensitivity level of IceCube, it is expected that the observed signal
could arise from the flux predicted here. A possible cutoff at high energies in this model varies with the redshift as $(1+z_0)^{-2/5}$ from the original spatial limitations for accelerating protons  (see Eq.\ \ref{emax_cr}). We predict in the observer frame
\begin{equation}
E_{\nu, max} \, = \, 10 \, {\rm PeV} \, \eta_{B, -1}^{-1/2} \, z_{1.3}^{-2/5} \, E_{57}^{2/5} \,.
\end{equation}

A measurement of the cutoff by IceCube would consequently help to constrain $\eta_{B} z_{1.3}^{4/5}$, the efficiency to run kinetic energy into magnetic field energy times the redshift to a power nearly unity, as the explosion energy is constrained by spatial limitations. The effect of   one remnant hitting the next remnant strongly reduces the overall emissivity.

The observed extragalactic gamma-ray flux exists at a
level of $E^{2}\cdot \left.dN/dE\right|_{\gamma}\approx
10^{-6}\,\diffunits$ at GeV energies (Abdo \etal
 2010). The flux predicted
here is at a level of $\sim 10^{-6.6}\,\diffunits$ at GeV energies. It
could therefore contribute to the total extragalactic flux  on the order of a 
few 10s of percent, depending on the exact values of the parameters adopted.  However, as is well understood, the gamma ray horizon is quite limited (e.g. Protheroe \& Biermann 1996; Dom{\'i}nguez \etal 2013), making such a detection impossible for any photon energies at high GeV.  On the other hand, at low GeV scales this may be possible.

Correspondingly it is possible to derive an inverse Compton X-ray background from these remnants, which may help to explain some small fraction of the observed background.  However, this contribution so strongly depends on parameters such as the spectral index, that we do not present that calculation here.

\subsection{Our Galaxy and other galaxies}
We can ask what flux in high energy neutrinos is expected from cosmic ray interactions in our own Galaxy (Berezinsky \etal 1993, Gaisser \etal 1995). Considering the two cosmic ray interaction sites identified (Stanev \etal 1993, Biermann \etal 2001, and Nath \etal
2012), in the interstellar medium, i) interactions mostly by cosmic ray protons with a spectrum close to $E^{-2.78}$, and ii) interactions in the massive shells of exploding Wolf-Rayet stars with a spectrum close to $E^{-2.33}$, at high energy we predict an upper cut-off of about $10^{14}$ eV for the flatter spectrum from the spectral turn-down at the cosmic ray knee, and a lower cut-off for the steeper spectrum. This spectrum refers to the polar cap component of the cosmic rays arising from massive stars exploding into their winds. Therefore most of the interactions can be expected to occur close to the sources, and so the spectrum may be quite close to the injection spectrum of $E^{-2}$.  The flux can readily be predicted to be of the same order of magnitude as observed, but a key difference is its distribution on the sky, which is obviously tightly correlated with the disk of the inner Galaxy and scales with the cosmic ray intensity in the part of the Galaxy observed.  Therefore in this picture any high energy neutrinos correlated with the inner Galaxy in direction are predicted to have a high energy spectrum of $E^{-2}$ and a cutoff of about 100 TeV. 
Considering Active Galactic Nuclei (AGN) we mention as an example of a summarizing paper Mannheim \etal ( 2001), which shows a proton-blazar model below the present data and an upper limit above the present data. Their Fig 3 also shows that the well-known simple Waxman \& Bahcall (1997) bound is rather close to the present data.
An AGN model matching the low radio counts and also explaining the neutrino background data remains to be worked out in detail.

There is a corresponding cosmic background from other galaxies. Since the far-infrared background is dominated by star-bursts and normal galaxies at redshifts of order unity (e.g. Lagache \etal 2005, Dole \etal 2006), we can expect an isotropic background due to these galaxies at such redshifts.  Scaling from the far-infrared this background can be estimated at a level which is about an order of magnitude below the background deduced above in the picture of black hole formation.  This is analogous to the finding that the radio background using known source populations is significantly lower (Condon \etal
2012), again by about an order of magnitude.  However, this background is predicted to reach only  neutrino energies lower than what has been seen, assuming that the knee energy of the cosmic ray spectrum is universal, although even then a matching flux would be difficult.

\section{Consequences and constraints}
\subsection{Galaxy formation}
Recently a very large space density of bulge-less disk galaxies has been detected, with total baryonic masses of order $10^{11} \; M_{\odot}$, and a space density of order $10^{-2}$  Mpc$^{-3}$ (Kormendy \etal 2010, 2011, Kormendy \& Bender 2011).  This corresponds to the mass and density of the gaseous massive shells formed around the explosions of the super-massive stars leading to the first generation of super-massive black holes. The gaseous mass covered by the shell as derived above is
\be
M_{shell} \; = \; 10^{10.4} \, {\rm M_{\odot}} \, E_{57}^{3/5} \, z_{1.3}^{-3/5} \, .
\ee
Going all the way to the next black hole gives the same mass at redshift 20, but a relatively higher mass at higher redshift. 
\be
M_{wall} \; = \; 10^{10.4} \, {\rm M_{\odot}} \, N_{BH,0,0}^{-1} \, ,
\ee
independent of redshift, and using the higher original black hole density suggested above of order $N_{BH,0,0} \simeq 1$.  The gaseous mass around a distribution of super-massive black holes will likely break into a fair number of pieces, but occasionally could also coagulate to make larger galaxies.  This process would allow a large number of galaxies of relatively large mass but which never merged, to exist, thus potentially solving the problem posed by Kormendy \etal
(2010), which is how to fit the observed existence of large bulge-less galaxies into a hierarchical formation scheme of galaxies.  

This also matches the discovery of a much larger space density of very massive disk-like galaxies at redshifts beyond 2 (Buitrago \etal
2013) than suggested by simulations.  At high redshifts massive galaxies have properties which are more disk-like in terms of their overall surface brightness profiles than similar mass galaxies today (e.g., Buitrago \etal 2013).  The visual morphologies suggest that a large fraction of these are disks in formation, and many of these systems could survive until today (Conselice \etal
2011).  This is especially the case if the formation of these galaxies is dominated by gas accretion, which appears to be the case (Conselice \etal
2013), rather than mergers which would fundamentally alter their structure into more spheroid like galaxies (Conselice 2013 priv.comm.).  Since the star formation time scales in disk-like galaxies are longer at greater distances, it could be that the galaxy disk just grows from the inside, while the outer gas very slowly settles into a star forming disk.

We also note that the merging between massive disk-like galaxies with other galaxies harboring black holes could produce the most massive black holes observed to date (e.g. Sbarrato \etal
 2013).

One can ask if we can possibly determine the redshift for the phase of formation of this population of supermassive black holes.
For the early universe molecular hydrogen is the best system to obtain a reliable redshift, and the lines of H D$^+$, H$_{2}$, H$_{2}^{+}$ or H$_{3}^{+}$ are clearly the most interesting (e.g.,  Nath \& Biermann 1994, Becker \etal
2011).  Using the column fraction relative to neutral hydrogen for certain transitions of $H_{3}^{+}$ found by Goto \etal (2008) of order $10^{-6.8}$,  and using the same column fraction relative to the total hydrogen,  the columns predicted for various level populations of $H_{3}^{+}$ might well approach or even exceed $10^{19}$  cm$^{-2}$, making a discovery a possibility. However, the very strong foreground would make such a determination very challenging.   

\subsection{Reionization}
The history of re-ionization (see, e.g., Coe \etal
2012) provides another check on these ideas.  The Thomson depth due to the radio remnants formed in the explosion of the super-massive stars, as well as that due to HII-regions accompanying the super-massive stars during their active life, can both be calculated.

The Thomson depth integral is
\begin{equation}
\int x_e(z') \, \sigma_T \, n_0 (1+z')^{3} \, \frac{c}{(1+z') H(z')} d z' \, .
\end{equation}
This is essentially an integral over the time a region remains ionized.
First we work out the Thomson depth due to the radio remnants, because their parameters have many more constraints due to our starting point of assuming that the radio background found by Fixsen \etal (2011), Kogut \etal (2011), and Seiffert \etal (2011), using the conditions determined by Condon \etal
 (2012), is real and can be explained by the radio remnants in the early universe.  This calculation uses the fact that most of the matter is in a narrow shell of $1/12$ of the radius, with a density increased by $4$ from strong shock conditions; a ray traverses a shell twice.  The Thomson depth is
\be
\tau_{Th, 1} \; = \; 10^{-5.3} \, E_{57}^{1/5} \, z_{1.3}^{+9/5} \, ,
\ee
using the integral above, giving a $(1+z)^{-1}$ correction.  The solid angle of a single remnant on the sky is given by
\be
10^{-7.9} \, E_{57}^{2/5} \, z_{1.3}^{-2/5} \, .
\ee

As noted earlier the total number of sources per solid angle is $10^{9.8}  \, N_{BH,0,0} \,z_{1.3}^{-1/2}$, and so the total solid angle subtended by all remnants per steradian is $10^{+1.9} \, N_{BH,0,0} \, E_{57}^{2/5} \, z_{1.3}^{-9/10}$.  Since this number exceeds unity, this means that along any single line of sight we have very many remnants.

Combining this overlap factor with the optical depth of a single remnant we get the overall optical depth
\be
\tau_{Th, \Sigma} \; = \; 10^{-3.4} \, N_{BH, 0, 0} \, E_{57}^{3/5}  \, z_{1.3}^{+9/10} \, .
\ee
This value is below the observations (Ade et al. 2013), which find a value of $\tau_{Th, \Sigma} \; \simeq \;  10^{-1.1}$. This discrepancy cannot be easily resolved, and shows that these HII regions do not contribute significantly to the integrated Thomson depth, unless both the redshift and the black hole density are both significantly higher than assumed here for reference.

There is also ionization by the very first stellar size black holes (Mirabel \etal
 2011).This ionization proceeds via the X-ray photons emitted as a result of accretion; X-ray photons are very effective in partially as well as fully ionizing gas over large distances.  These pathways to ionization may well dominate over the effect of the radio remnants formed as a result of the first generation of super-massive black holes. However, they distinguish themselves by varying over a large redshift range from early on, whereas the effect caused by the formation of the first generation of super-massive black holes can occur only at such high redshifts that stellar winds have not yet formed, that is only at extremely low heavy element abundances.

The corresponding Thomson depth due to the HII regions of the super-massive stars preceding the formation of the black holes can also be worked out, but such a calculation suffers from the extreme uncertainty that we cannot be sure that the very outer layers of the growing super-massive star is sufficiently close to hydrostatic equilibrium to actually radiate massively in the ultraviolet.  Whatever small fraction emerges as ultraviolet to ionize Hydrogen is a fraction $\epsilon$ of the Eddington luminosity. Such an analysis is extremely uncertain, and we do not reproduce it here, but do note one limit, if the emission is shifted into the far infrared.

Much of this emission could be shifted to the red, which for the observer is the infrared.  This phase of the evolution could produce some infrared, which limits what is possible.  Again, we can calculate the maximal contribution to the IR background
\begin{eqnarray}
F_{IR, max}  &=&  10^{-6.0} \,   N_{BH,0,0} \, M_{SMS, 6.5} \, z_{1.3}^{-1} \nonumber\\
&& {\rm erg \, cm^{-2} \, s^{-1} \, sr^{-1} }\,,
\end{eqnarray}
and use the full Eddington luminosity to obtain a safe constraint. This is within an order of magnitude of the observed, very uncertain, minimum level around 10 $\mu$ of the IR background (Lagache \etal 2005, Dole \etal 2006).  In fact, using the increased density of early black holes, the luminosity may come close to this limit.  However, we have to emphasize that the IR and FIR background are readily explained by the well known galaxy population at redshift of order unity, and so any further contribution must be small.  On the other hand, the source density of the population discussed in this paper has a much higher density on the sky than the known galaxy populations (Condon \etal 2012).

\subsection{Constraints on the model}
Taken together we now have several constraints with rather similar combined dependencies on explosion energy and redshift.  These constraints include the radio background, the individual source flux density, and Thomson depth of the remnants, all of which involve some power of the explosion energy close to unity, 
multiplied by the redshift (in units of $(1+20)$) to another power also  close to unity.  

Matching the radio background  suggests an increase of the original black hole density to about $N_{BH,0} \simge 1 \, Mpc^{-3}$.  This leads to an approximate match of the Thomson depth.  The remaining missing factor may be best accommodated by a higher redshift, possibly up to about 50, still consistent with the original concept of very early massive star formation.

The Condon \etal (2012) flux density limit gives an upper bound for a combination of the individual explosion energy and the redshift.  The Condon \etal (2012) number limit gives a lower limit for the original black hole density.  However, the use of such limits is somewhat dubious, since the sources are extended beyond the Condon \etal (2012) beam, and very large overlaps between sources are present.


The space constraint derives from the explosion energy: if we wish to have the Hubble-time limited remnants remain below the space constraint in order to allow sufficient time for the remnants to radiate, then the explosion energy can certainly not be significantly larger than our nominal value $10^{57}$ ergs. Of course it might be lower, at the price of forcing us to a higher redshift to match the observed backgrounds.

The Thomson depth using the remnants gives a reasonable match. 

\subsection{Other tests}
Other tests involve the production and distribution of magnetic fields, the distribution of heavy elements (e.g. Simcoe \etal  2012) and further consequences from cosmic ray interactions.  An especially interesting question is whether the large number of x-ray photons and cosmic rays at low energies could precipitate the formation of abundant molecular Hydrogen including deuterated forms and thus allow the formation of a larger initially bound system consisting of many smaller systems with a million Solar mass black hole. This could possibly allow,  through a second level gravo-thermal collapse, the early formation of dense galaxies with extreme black holes in regions of higher general density, which would later turn into clusters such as Perseus or Virgo (van den Bosch \etal 2012).

\section{Conclusions}
We have derived the radio background due to the formation of the first population of super-massive black holes.  Their production is assumed to lead to radio remnants, quite similar to normal supernova remnants, just scaled up.

The prediction presented here falls within the IceCube sensitivity (Schukraft \etal  2013) and therefore provides a possible explanation of the recently announced IceCube neutrino excess (Aartsen \etal
2013).   The model also leads to a possible explanation of the observed flux density and spectrum of the gamma-ray background.   

The model obeys all known radio observational constraints, including single source strength, total number, lack of far-infrared emission, and radio spectrum.  Adopting cautious values for the parameters of the model suggests that the formation redshift may be quite large, consistent with 
a very early epoch of star formation.
Matching both the neutrino and radio background gives
rather strong constraints on the factor $\eta_{\rm B} ^{0.8} \eta_{\rm CR,e}/\eta_{\rm CR}$, since the 
ratio of these two backgrounds depends only weakly on the explosion energy and redshift.
If upon further exploration the radio background is lowered in its flux density (e.g.,
see Subrahmanyan \& Cowsik 2013), then the explanation of the neutrino background stands, and this model would then predict a radio background at a level depending on this factor, relative to the neutrino background.

Interestingly, the scenario also has the potential to solve the formation riddle of large massive galaxies, which to all appearances never merged (see Kormendy \etal
2010). The massive shells formed by the explosion of the super-massive stars give the right order of magnitude both for the mass and the space density to form such galaxies.  A direct consequence is that a large number of galaxies never merged.
The redshift in this scenario could possibly be determined using the absorption spectra of hydrogen molecules, H$_{2}$, HD$^{+}$, H$_{2}^{+}$ or H$_{3}^{+}$, although the observations could be challenging.

If the interpretation can be confirmed, it would demonstrate the formation of the first super-massive black holes in the universe.  The radio emission is non-thermal and together with the recent detection of a high energy neutrino background constitutes evidence for the first cosmic ray population in the universe.  

\section{Acknowledgements}
Discussions of this topic and related issues with R. Buta, Ch. Conselice, T. En{\ss}lin, W. Keel, T. Kneiske, A. Kogut, J. Kormendy, M. Kramer, A. Lasenby, K. Menten, F. Mirabel, S. Paduroiu, J. Rachen, N. Sanchez, E.-S. Seo, and H. de Vega are gratefully acknowledged, and the comments of an anonymous referee is also greatly appreciated.   L.I.C. is partially supported by CNCSIS Contract 539/2009 and CNMP Contract 82077/2008. L.I.C. is member of the International Max Planck Research School (IMPRS) for Astronomy and Astrophysics at the Universities of Bonn and Cologne.  The work of B.C.H. was supported in part from DOE grant DE-FG02-10ER41714.  The key prediction of a neutrino background has been presented in an invited lecture at a Helmholtz-Alliance meeting in Berlin-Zeuthen Sep 19, 2012; this Alliance includes the German IceCube team:
https://indico.desy.de/conferenceDisplay.py?confId=5709.


\begin{thebibliography}{999}


\bibitem[\protect\citeauthoryear{Aartsen}%
{2013}]{aartsen13}  Aartsen, M. G., \etal (IceCube-Coll.) 2013, \Science, 342, 1242856



\bibitem[\protect\citeauthoryear{Abbasi}%
{2011}]{abbasi11}  Abbasi, R., \etal (IceCube-Coll.) 2011, \PRD , 84, 082001

\bibitem[\protect\citeauthoryear{Abdo}%
{2010}]{abdo10}  Abdo, A. A., \etal, (Fermi-LAT Coll.) 2010 \PRL ,104, 101101

\bibitem[{ade}{2013}]{ade13}   Ade, P.A.R., \etal (PLANCK collaboration) 2013, eprint arXiv:1303.5076


\bibitem[\protect\citeauthoryear{Amaro}%
{2013}]{amaro13}Amaro-Seoane, P., Konstantinidis, S., Dewi Freitag, M., Coleman, M., Rasio, F. A. 2012, eprint arXiv:1211.6738


\bibitem[\protect\citeauthoryear{Appenzeller1}%
{1972}]{appenzeler72} Appenzeller, I., \&  Fricke, K. 1972a, \AA,18, 10  

\bibitem[\protect\citeauthoryear{Appenzeller2}%
{1972}]{appenzeler72b}  Appenzeller, I., \&  Fricke, K. 1972b, \AA, 21, 285

\bibitem[\protect\citeauthoryear{Becker}%
{2010}]{becker10}  Becker, J. K., \& Biermann, P. L. 2009, \ApP, 31, 138
	
\bibitem[\protect\citeauthoryear{Becker}%
{2011}]{becker11} Becker, J.K., \etal 2011, \ApJL, 739, 43 
	
\bibitem[\protect\citeauthoryear{Begelman}%
{2006}]{begelman06}  Begelman, M.C., Volonteri, M., Rees, M.J. 2006, \MNRAS, 370, 289

\bibitem[\protect\citeauthoryear{Bell}%
{2001}]{bell01} Bell, A. R., \& Lucek, S. G. 2001, \MNRAS, 321, 433


\bibitem[\protect\citeauthoryear{Bellovary}%
{2011}]{bellovary11} Bellovary, J., Volonteri, M., Governato, F., Shen, S., Quinn, T., Wadsley, J. 2011, \ApJ, 742, 13 

\bibitem[\protect\citeauthoryear{Berezinsky}%
{1993}]{berezinsky93}  Berezinsky, V. S., Gaisser, T. K., Halzen, F., Stanev, T 1993, \ApP, 1, 281

\bibitem[\protect\citeauthoryear{Berezinsky}%
{2012}]{berezinsky12} Berezinsky, V., \& Blasi, P. 2012, \PRD, 85, 123003


	
\bibitem[\protect\citeauthoryear{biermann1}%
{1993}]{biermann93} Biermann, P.L. 1993, \AA, 271, 649 

\bibitem[\protect\citeauthoryear{Biermann}%
{1993}]{biermann93b} Biermann, P.L., \& Cassinelli, J.P. 1993, \AA, 277,
  	691 

\bibitem[\protect\citeauthoryear{Biermann}%
{1993}]{biermann93c} Biermann, P.L., \& Strom, R.G. 1993, \AA, 275, 659 

\bibitem[\protect\citeauthoryear{Biermann}%
{2006}]{biermann06} Biermann, P.L., \& Kusenko, A. 2006, \PRL, 96, 091301 

\bibitem[\protect\citeauthoryear{Biermann}%
{2012}]{biermann12} Biermann, P.L., \& Harms, B. C. 2012, eprint arXiv:1205.4016 

\bibitem[\protect\citeauthoryear{Biermann}%
{2012}]{biermann12b} Biermann, P.L., \&  Harms, B.C. 2012, in {it Proc. at the 13th Marcel Grossmann
Meeting}, July 2012, eprint arXiv:1302.0040

\bibitem[\protect\citeauthoryear{Bishnovatyi-kogan}%
{1973}]{bisnovatyi-kogan73} Bisnovatyi-Kogan, G. S., Ruzmaikin, A. A., \& Syunyaev, R. A. 1973, {\it Sov. Astr.}, 17, 137%

\bibitem[\protect\citeauthoryear{Bonoli}%
{2013}]{bonoli13} Bonoli, S., Mayer, L., Callegari, S. 2013, \MNRAS, in press (eprint arxiv/1211.3752)

\bibitem[\protect\citeauthoryear{vandenBosch}%
{2012}]{vandenbosch12}  van den Bosch, R.C.E., Remco C. E., Gebhardt, K., G�ltekin, K.,
van de Ven, G., van der Wel, A., Walsh, J. L. 2012, \Nature, 491, 729


	
\bibitem[\protect\citeauthoryear{Buitrago}%
{2013}]{buitrago13} Buitrago, F., Conselice, C. J.,Epinat, B.,
Bedregal, A. G., Grutzbauch, R. 2013, eprint arXiv:1305.0268 

\bibitem[\protect\citeauthoryear{Caramete}%
{2010}]{caramete10}  Caramete, L.I., \& Biermann, P.L.2010 , \AA, 521, 55 

\bibitem[\protect\citeauthoryear{Chandrasekhar}%
{1939}]{chandrasekhar39} Chandrasekhar, S., {\it An introduction to the study of stellar structure}, 
Chicago, Ill., The University of Chicago press (1939)


\bibitem[\protect\citeauthoryear{Condon}%
{2012}]{condon12} Condon, J.J., Cotton, W. D., Fomalont, E. B., Kellermann, K. I., Miller, N., Perley, R. A., Scott, D., Vernstrom, T., Wall, J. V. 2012, \ApJ, 758, 23 

\bibitem[\protect\citeauthoryear{Conselice}%
{2011}]{conselice11} Conselice, C. J., Bluck, A. F. L., Ravindranath, S., Mortlock, A.,
Koekemoer, A. M., Buitrago, F., Gr�tzbauch, R., Penny, S. J. 2011, \MNRAS, 417, 2770 

\bibitem[\protect\citeauthoryear{Conselice}%
{2013}]{conselice13}  Conselice, C. J., Mortlock, A., Bluck, A. F. L.,
Gr�tzbauch, R., Duncan, K. 2013, \MNRAS, 430, 1051

\bibitem[\protect\citeauthoryear{Cox}%
{1972}]{cox72} Cox, D. P. 1972, \ApJ,178, 159 
	

\bibitem[\protect\citeauthoryear{Destri}%
{2012}]{destri12}  Destri, C., de Vega, H. J., \& Sanchez, N. G. 2012, {\it NewA}, 22, 39 
	
\bibitem[\protect\citeauthoryear{Dole}%
{2006}]{dole06} Dole, H., Lagache, G., Puget, J. 2006, {\it The Spitzer Space Telescope: New Views of the Cosmos}, . Eds. L. Armus \& W.T. Reach.  ASP Conf. Ser., 357, 290 

\bibitem[\protect\citeauthoryear{Dominguez}%
{2013}]{dominguez13}  Dom{\'i}nguez, A., Finke, J. D., Prada, F., Primack, J. R., Kitaura, F. S., 
Siana, B., Paneque, D. 2013, \ApJ, 770,  77 


\bibitem[\protect\citeauthoryear{Drury}%
{1983}]{drury83}  Drury, L. O'C. 1983, \RPP, 46, 973 
	
\bibitem[\protect\citeauthoryear{Everett}%
{2010}]{everett10}  Everett, J.E., Schiller, Q. G., Zweibel, E. G. 2010, \ApJ, 711, 13 

	

\bibitem[\protect\citeauthoryear{Formengo}%
{2014}]{formengo14}  Formengo, N., Roberto, A., Lineros, M. R., Taoso, M. 2014, preprint, arXiv:1402.2218

\bibitem[\protect\citeauthoryear{Fixsen}%
{2011}]{fixsen11}  Fixsen, D.J., \etal 2011, \ApJ, 734, 5 



\bibitem[\protect\citeauthoryear{Gaisser}%
{1995}]{gaisser95} Gaisser, T. K., Halzen, F., Stanev, T. 1995, {\it Phys. Rep.}, 258, 173 

\bibitem[\protect\citeauthoryear{Geach}%
{2013}]{geach13} Geach, J.E., \etal 2013, \MNRAS, 432, 53 

\bibitem[\protect\citeauthoryear{Gargely}%
{2009}]{gargely09} Gergely L., Biermann, P.L. 2009, \ApJ, 697, 1621 

\bibitem[\protect\citeauthoryear{Goto}%
{2008}]{goto08}  Goto, M., \etal 2008, \ApJ, 688, 306 

\bibitem[\protect\citeauthoryear{Greene}%
{2006}]{greene06} Greene, J.E., Barth, A.J., \& Ho, L.C., 2006, \NewAR, 50, 739 

\bibitem[\protect\citeauthoryear{Greene}%
{2007}]{greene07a}  Greene, J.E., \& Ho, L.C. 2007a, \ApJ, 670, 92 

\bibitem[\protect\citeauthoryear{Greene}%
{2007}]{greene07b} Greene, J. E., \&  Ho, L. C. 2007b, in proc. {\it The Central Engine of Active Galactic Nuclei}, ASP Conf., 373, 33
	
\bibitem[\protect\citeauthoryear{Greene}%
{2008}]{greene08} Greene, J.E., Ho, L.C., \& Barth, A.J. 2008, \ApJ, 688, 159 

	

\bibitem[\protect\citeauthoryear{Hoflich}%
{1999}]{hoflich99} H{\"o}flich, P., Wheeler, J. C., \& Wang, L. 1999, \ApJ, 521, 179 


\bibitem[\protect\citeauthoryear{Holder}%
{2004}]{holder04} Holder, G. P. 2014, ApJ, 780, 112

\bibitem[\protect\citeauthoryear{Jarosik}%
{2011}]{jarosik11} Jarosik, N. \etal 2011, \ApJS, 192, 14 

\bibitem[\protect\citeauthoryear{Jokipii}%
{1977}]{jokipii77} Jokipii, J.R., Levy, W.H., Hubbard, E.B. 1977, \ApJ, 213,  861 

\bibitem[\protect\citeauthoryear{Jokipii}%
{1987}]{jokipii87} Jokipii, J.R. 1987, \ApJ, 313, 842  




\bibitem[\protect\citeauthoryear{Kogut}%
{2003}]{kogut03} Kogut, A., \etal 2003, \ApJS, 148, 161 

\bibitem[\protect\citeauthoryear{kogut}%
{2011}]{kogut11} Kogut, A., \etal 2011, \ApJ, 734, 4 

	
\bibitem[\protect\citeauthoryear{Kormendy}%
{2010}]{kormendy10} Kormendy, J., Drory, N., Bender, R., Cornell, M. E. 2010, \ApJ, 723, 54 
	
\bibitem[\protect\citeauthoryear{Kormendy}%
{2011}]{kormendy11} Kormendy, J., Bender, R., \& Cornell, M. E. 2011, \Nature, 469, 374 

\bibitem[\protect\citeauthoryear{Kormendy}%
{2011}]{kormendy11b}  Kormendy, J., Bender, R. 2011, \Nature, 469, 377 
	
	
\bibitem[\protect\citeauthoryear{Legache}%
{2005}]{lagache05}  Lagache, G., Puget, J.-L., Dole, H. 2005, \ARAA, 43, 727 

\bibitem[\protect\citeauthoryear{Lucek}%
{2000}]{lucek00} Lucek, S. G., \& Bell, A. R. 2000, \MNRAS, 314, 65 

\bibitem[\protect\citeauthoryear{Mannheim}%
{2001}]{mannheim01} Mannheim, K., Protheroe, R. J., Rache, J. P. 2001, \PRD, 63, 023003
	 
\bibitem[\protect\citeauthoryear{McMillan}%
{2007}]{mcmillan07} McMillan, S. L. W., Portegies Zwart, S. F. 2007, in Proc. {\it Massive Stars in Interactive Binaries}, ASP Conf. Ser. 367. Eds. N. St.-Louis \& A. F.J. Moffat. San Francisco: A.S.P., 697

\bibitem[\protect\citeauthoryear{meiksin}%
{2013}]{meiksin13} Meiksin, A., Whalen, D. J. 2013, MNRAS, 430, 2854

\bibitem[\protect\citeauthoryear{Mirabel}%
{2011}]{mirabel11}  Mirabel, I.F., Dijkstra, M., Laurent, P., Loeb, A., Pritchard, J. R. 2011, \AA, 528, 149  
	

\bibitem[\protect\citeauthoryear{Mortlock}%
{2011}]{mortlock11} Mortlock, D.J., \etal 2011, \Nature, 474, 616 

\bibitem[\protect\citeauthoryear{Munyaneza}%
{2005}]{munyaneza05}  Munyaneza, F., \& Biermann, P. L. 2005, \AA, 436, 805 
	
\bibitem[\protect\citeauthoryear{Munyaneza}%
{2006}]{munyaneza13} Munyaneza, F., \& Biermann, P. L. 2006, \AAL, 458, L9 
	
\bibitem[\protect\citeauthoryear{Nakamura}%
{2001}]{nakamura01} Nakamura, T., Mazzali, P. A., Nomoto, K., Iwamoto, K. 2001, \ApJ, 550, 991 
	
\bibitem[\protect\citeauthoryear{Nath}%
{1994}]{nath94} Nath, B.B. \& Biermann, P.L. 1994, \MNRAS, 267, 447 



\bibitem[\protect\citeauthoryear{Zwart}%
{2004}]{zwart13} Portegies Zwart, S. F., Baumgardt, H., Hut, P., Makino, J., McMillan, St. L. W. 2004, \Nature, 428, 
724 

\bibitem[\protect\citeauthoryear{Zwart}%
{2007}]{zwart107}  Portegies Zwart, S. F., \& van den Heuvel, E. P. J. 2007 , 
\Nature,  450, 388 

\bibitem[\protect\citeauthoryear{Zwart}%
{2010}]{zwart10}  Portegies Zwart, S. F., McMillan, St. L. W., \& Gieles, M. 2010, \ARAA, 48, 431 


\bibitem[\protect\citeauthoryear{Protheroe}%
{1996}]{protheroe96} Protheroe, R.J., \& Biermann, P.L. 1996, \ApP, 6, 45 

\bibitem[\protect\citeauthoryear{Pugliese}%
{2000}]{pugliese00} Pugliese, G., Falcke, H., Wang, Y.-P., \& Biermann, P.-L. 2000, \AA, 358, 409 

\bibitem[\protect\citeauthoryear{Quinlan}%
{1990}]{quinlan90} Quinlan, G. D., \& Shapiro, S. L. 1990, \ApJ, 356, 483 

\bibitem[\protect\citeauthoryear{Ruhe}%
{2013}]{ruhe13} Ruhe, T. (IceCube-Coll) \etal 2013, ICRC 2013, Brazil

\bibitem[\protect\citeauthoryear{Sanders}%
{1970}]{sanders70}  Sanders, R. H. 1970, \ApJ, 162, 791 

\bibitem[\protect\citeauthoryear{Sbarrato}%
{2013}]{sbarrato13}	Sbarrato, T., Ghisellini, G., Nardini, M., Tagliaferri, G., Greiner, J., Rau, A., Schady, P. 2013, eprint  arXiv:1303.6951 

\bibitem[\protect\citeauthoryear{Schukraft}%
{2013}]{schukraft13} Schukraft, A. (for the IceCube-Coll) 2013, {\it Nucl. Phys. B (Proc. Suppl.)}, 237, 266 

	
	
\bibitem[\protect\citeauthoryear{Seiffert}%
{2011}]{seiffert11} Seiffert, M., \etal 2011, \ApJ, 734, 6 
	
\bibitem[\protect\citeauthoryear{Silk}%
{1998}]{silk98} Silk, J., Rees, M. J. 1998, \AAL,  331, L1 

\bibitem[\protect\citeauthoryear{Simcoe}%
{2012}]{simcoe11} Simcoe, R.A., Sullivan, P. W., Cooksey, K. L., Kao, M. M., Matejek, M. S., Burgasser, A. J. 2012, \Nature, 492, 79
    
\bibitem[\protect\citeauthoryear{Singal}%
{2010}]{singal10} Singal, J., Stawarz, L., Lawrence, A., Petrosian, V. 2010, MNRAS, 409, 1172
    
\bibitem[\protect\citeauthoryear{Spitzer}%
{1969}]{spitzer69} Spitzer, L., Jr. 1969, \ApJL, 158, L139 

\bibitem[\protect\citeauthoryear{Spitzer}%
{1987}]{spitzer87}  Spitzer, L., Jr. 1987, {\it Dynamical evolution of globular clusters}, Princeton Univ. Press


\bibitem[\protect\citeauthoryear{Stern}%
{2012}]{stern12} Stern, D., \etal 2012, \ApJ, 753, 30 



\bibitem[\protect\citeauthoryear{Subrahmanyan}%
{2013}]{subrahmanyan13} Subrahmanyan, R., Cowsik, R. 2013, \ApJ, 776, 42


\bibitem[\protect\citeauthoryear{Sun}%
{2008}]{sun08} Sun, X. H., Reich, W., Waelkens, A., En\ss lin, T. A. 2008, \AA, 477, 573 



\bibitem[\protect\citeauthoryear{Voit}%
{1996}]{voit96} Voit, G. M. 1996, \ApJ, 465, 548 

\bibitem[\protect\citeauthoryear{Wang}%
{1998}]{wang98}  Wang, Y.-P., \& Biermann, P.L. 1998, \AA, 334, 87 


\bibitem[\protect\citeauthoryear{Waxman}%
{1997}]{waxman97}  Waxman, E., Bahcall, J1997, \PRL 78, 2292

\bibitem[\protect\citeauthoryear{Weibel}%
{1959}]{weibel59}  Weibel, E. S. 1959, \PRL, 2, 83 

\bibitem[\protect\citeauthoryear{Weiler}%
{2002}]{weiler02} Weiler, K. W., Panagia, N., Montes, M., Marcos, J., Sramek, R. A. 2002, \ARAA, 40, 387 

\bibitem[\protect\citeauthoryear{Whalen}%
{2013}]{whalen13} Whalen, D. J. \etal 2013, \ApJ, 778, 17


\bibitem[\protect\citeauthoryear{Woosley}%
{2002}]{woosley02} Woosley, S. E., Heger, A., \& Weaver, T. A. 2002, \RMP, 74, 1015 

\bibitem[\protect\citeauthoryear{Yungelson}%
{2008}]{yungelson08} Yungelson, L.R., van den Heugel, E. P. J., Vink, J. S., Portegies Zwart, S. F., de Koter, A. 2008, \AA, 477, 223 

\bibitem[\protect\citeauthoryear{Zinn}%
{2011}]{zinn11} Zinn, P.-C., Middelberg, E., \& Ibar, E. 2011, \AA, 531, 14 

\end{thebibliography}
\end{document}